\documentclass[11pt]{article}

\usepackage[margin=1.25in]{geometry}

\usepackage{amsmath}
\usepackage{amsfonts}
\usepackage{amsthm}
\usepackage{graphicx}

\usepackage[T1]{fontenc}
\usepackage[default]{lato}

\usepackage{microtype}

\usepackage{tikz}

\usepackage{xcolor}

\definecolor{dark-maroon}{HTML}{5D0F0D}
\definecolor{navyblue}{HTML}{0A3044}

\definecolor{purple}{HTML}{5601A4}
\definecolor{navy}{HTML}{0D3D56}
\definecolor{ruby}{HTML}{9a2515}
\definecolor{alice}{HTML}{107895}
\definecolor{daisy}{HTML}{EBC944}
\definecolor{coral}{HTML}{F26D21}
\definecolor{kelly}{HTML}{829356}
\definecolor{cranberry}{HTML}{E64173}
\definecolor{jet}{HTML}{131516}
\definecolor{asher}{HTML}{555F61}
\definecolor{slate}{HTML}{314F4F}

\usepackage{hyperref}
\hypersetup{
    colorlinks= true,
    citecolor= dark-maroon,
    linkcolor= dark-maroon,
    filecolor= dark-maroon,      
    urlcolor= dark-maroon,
}

\usepackage{natbib}
\setcitestyle{comma,aysep={}}

\newtheoremstyle{spacing}
{}
{}
{}
{}
{\bfseries\color{navyblue}}
{.\ }
{2.5mm}
{}

\theoremstyle{spacing}

\newtheorem{proposition}{Proposition}
\newtheorem{assumption}{Assumption}

\global\long\def\expec#1{\mathbb{E}\left[#1\right]}%
\newcommand{\condexpec}[2]{\mathbb{E}\left[#1 \ \vert \ #2\right]}
\global\long\def\one{\mathbf{1}}%

\usepackage{titling}
\usepackage{setspace}

\pretitle{\begin{spacing}{1}\begin{flushleft}\huge}
\posttitle{\end{flushleft}\end{spacing}\vspace{-5mm}}
\preauthor{\begin{flushleft}\LARGE}
\postauthor{\end{flushleft}\vspace{-7.5mm}}
\predate{\begin{flushleft}\scshape\Large\color{asher}}
\postdate{\end{flushleft}\vspace{-5mm}}

\renewenvironment{abstract}
 {\noindent\rule{\linewidth}{.5pt}\noindent}
 {\noindent\rule{\linewidth}{.5pt}}

\usepackage[explicit]{titlesec}

\titleformat{\section}
  {\Large \bf \color{navyblue}}
  {\thesection \,---}
  {0.25em}
  {#1}
  
\titleformat{\subsection}
  {\fontsize{11}{10}\it}
  {\thesubsection.}
  {1em}
  {#1}

\let\oldfootnote\footnote
\renewcommand\footnote[1]{\oldfootnote{\ #1}}

\makeatletter
\renewcommand\@makefntext[1]{%
    \parindent 1em \noindent
    \hb@xt@1.8em{\hss\normalfont\@thefnmark.\hfill}#1
  }
\makeatother

\usepackage{enumitem}
\setitemize{labelindent=0.5em,labelsep=0.25cm,leftmargin=*}
\setenumerate{labelindent=0.5em,labelsep=0.25cm,leftmargin=*}

\usepackage{caption}

\DeclareCaptionLabelSeparator{threedash}{\,---\,}
\DeclareCaptionFont{navyblue}{\color{navyblue}}
\DeclareCaptionFont{jet}{\color{jet}}
\usepackage{subcaption}
\makeatletter
\let\input\@@input
\makeatother

\usepackage{adjustbox}

\usepackage{array}

\usepackage[flushleft]{threeparttable}
\usepackage{booktabs}

\usepackage{tabularx}
\newcolumntype{L}{>{\raggedright\arraybackslash}X}
\newcolumntype{R}{>{\raggedleft\arraybackslash}X}
\newcolumntype{C}{>{\centering\arraybackslash}X}

\usepackage{pdflscape}

\usepackage{fancyhdr}
\fancypagestyle{lscaped}{%
    \fancyhf{}
    
    \textnormal
    \fancyfoot{%
        \tikz[remember picture,overlay]
        \node[outer sep=2.5cm,above,rotate=90] at (current page.east) {\thepage};
    }
}

\title{Difference-in-Differences with Geocoded Microdata\thanks{I am grateful to Taylor Jaworski, Damian Clarke, Alexander Bentz, James Flynn, Brach Champion, and Hannah Denker for the helpful insights.}}
\author{\href{https://kylebutts.com/}{Kyle Butts}\thanks{University of Colorado, Boulder. Email: \href{mailto:kyle.butts@colorado.edu}{kyle.butts@colorado.edu}.} 
}
\date{\today}

\newcommand{\dist}{\text{Dist}}

\hypersetup{pdftitle={Difference-in-Differences with Geocoded Microdata}, pdfauthor={Kyle Butts}}

\begin{document}

\begin{titlepage}
\maketitle

\begin{abstract}
    This paper formalizes a common approach for estimating effects of treatment at a specific location using geocoded microdata. This estimator compares units immediately next to treatment (an inner-ring) to units just slightly further away (an outer-ring). I introduce intuitive assumptions needed to identify the average treatment effect among the affected units and illustrates pitfalls that occur when these assumptions fail. Since one of these assumptions requires knowledge of exactly how far treatment effects are experienced, I propose a new method that relaxes this assumption and allows for nonparametric estimation using partitioning-based least squares developed in \citet{Cattaneo_Crump_Farrell_Feng_2019,Cattaneo_Farrell_Feng_2019}. Since treatment effects typically decay/change over distance, this estimator improves analysis by estimating a \emph{treatment effect curve} as a function of distance from treatment. This is contrast to the traditional method which, \emph{at best}, identifies the average effect of treatment. To illustrate the advantages of this method, I show that \citet{Linden_Rockoff_2008} under estimate the effects of increased crime risk on home values closest to the treatment and overestimate how far the effects extend by selecting a treatment ring that is too wide.

    \par~\par\noindent
    {\color{asher}JEL-Classification:} C13, C14, C18
    \par\noindent
    {\color{asher}Keywords:} Spatial Econometrics, Difference-in-Differences, Nonparametric Estimation
    \par\vspace{-2.5mm}
\end{abstract}
\end{titlepage}


\section{Introduction}

The rise of microdata with precisely geocoded locations has allowed researchers to begin answering questions about the effects of spatially-targeted treatments at a very granular level. What are the effects of local pollutants on child health?\footnote{See, e.g., \citet{Currie_Davis_Greenstone_Walker_2015} and \citet{Marcus_2021}.} Does living within walking distance to a new bus stop improve labor market outcomes?\footnote{See, e.g., \citet{Gibbons_Machin_2005} and \citet{Billings_2011}.} How far do neighborhood shocks, such as foreclosures or new construction spread?\footnote{See, e.g., \citet{Asquith_Mast_Reed_2021,Cui_Walsh_2015,Gerardi_Rosenblatt_Willen_Yao_2015} and \citet{Campbell_Giglio_Pathak_2011}.} When treatment is located at a specific point in space, a standard method of evaluating the effects of the treatment is to compare units that are close to treatment to those slightly further away -- what I will label the `ring method'. This paper formalizes the assumptions required for identification in the ring method, highlighting potential pitfalls of the currently used estimator, and proposes an improved estimator which relaxes these assumptions. 

\begin{figure}[tb]
    \caption{Rings Method}
    \label{fig:example-id}

    \begin{adjustbox}{width = 0.4\textwidth, center}
        \includegraphics[width=\textwidth]{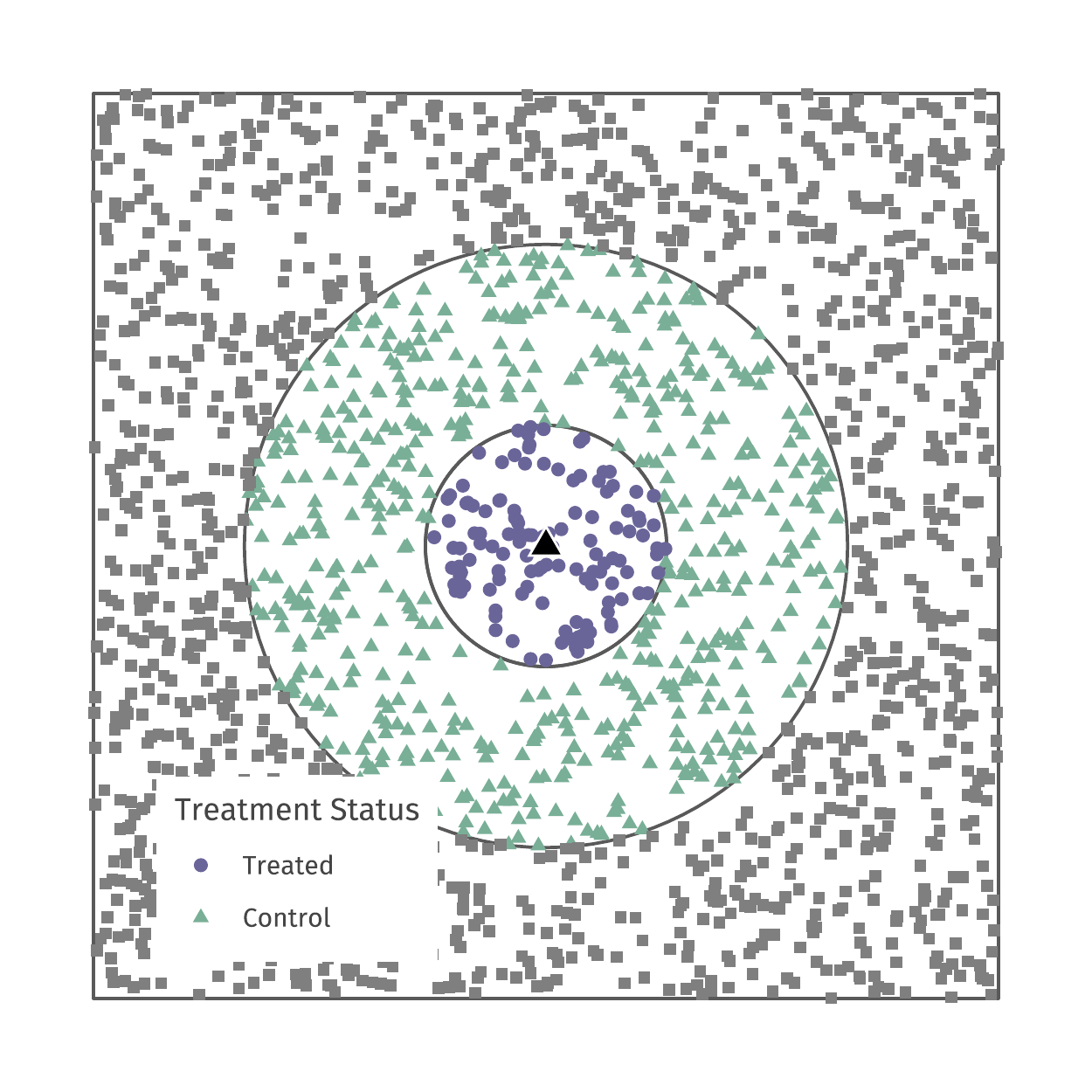}
    \end{adjustbox}
\end{figure}

The ring method is illustrated in \autoref{fig:example-id}. The center of the figure is marked with a triangle which represents the location of treatment, e.g. a foreclosed home. Units within the inner circle, marked by dots, are considered treated due to their proximity to the treatment location; units between the inner and outer circles, marked in triangles, are considered control units; and then the remaining units are removed from the sample. The appeal of this identification strategy is that since the treated and control units are all very close in physical location, e.g. having access to the same labor market and consumptive amenities, the counterfactual untreated outcomes will approximately be equal for units within each ring. The ring estimate for the treatment effect compares average changes in outcomes between units in the inner `treated' ring and the outer `control' ring to form an estimate for the treatment effect, i.e. a difference-in-differences estimator. 

My first contribution is to fill a gap in the econometrics literature by formalizing the necessary assumptions for unbiased estimates of the average treatment effect on the affected units.\footnote{This generalizes the treatment effect on the treated in the case where treatment isn't assigned to specific units.} The first assumption is the well understood parallel trends assumption for the treated and control units. This requires that the \emph{average} change in (counterfactual) untreated outcomes in the treated ring is equal to the \emph{average} change in the control ring. This allows the control units to estimate the counterfactual trend for the treated units. 

The second assumption requires the researcher to correctly identify how far treatment effects are experienced (the inner ring). This is a very strict assumption that when not satisfied, causes biased estimates of the treatment effect. If the treated ring is too narrow, then units in the control ring experience effects of treatment and the change among `control' units would no longer identify the counterfactual trend. On the other hand, if the treated ring is too wide, then the zero treatment effect of some unaffected units are averaged into the change among `treated' units. Therefore, results will be attenuated towards zero.


Since researchers often do not know how far treatment effects extend in most circumstances, I propose an estimator that replaces the second assumption with a less strict assumption by using a nonparametric, partitioning-based, least square estimator \citep{Cattaneo_Crump_Farrell_Feng_2019,Cattaneo_Farrell_Feng_2019}. My proposed methodology estimates the treatment effect curve as a function of distance by using many rings rather than trying to estimate the average treatment effect with one inner ring. This method requires that treatment effects become zero somewhere between the distance of 0 and the control ring \emph{without} the need to specify the exact distance. It however requires a stronger assumption that the counterfactual trend is constant across distance.\footnote{Note that the original requirement is that the \emph{average} change in each ring is equal but allows variation across distance.} This new assumption is more strict in that the standard method only requires that parallel trends holds \emph{on average} in each ring. However, researchers motivate the identification strategy by saying within a small distance from treatment that units are subject to a common set of shocks which implies the more strict assumption. While this assumption is not directly testable, the estimator creates a set of point estimates of treatment effects that can be used to visually inspect the plausability of the assumption. If after some distance, treatment effects become centered at zero, this suggests that common trends hold, akin to the pre-trends test in event study regressions.

The nonparametric approach allows the researcher to get a more complete picture of how the intervention affects units at various distances rather than estimating an ``overall effect''. For example, the construction of a new bus-stop potentially creates net costs to immediate neighbors while providing net benefits for homes slightly further away. Estimation of the treatment effect curve can illustrate these different effects that the ``overall effect'' would mask. In this case, the average effect could be zero even though most units experience non-zero effects.

\subsection{Relation to Literature}

This paper relates to a few papers that address difficulties with using the rings method for causal effect estimation. In the online appendix, \citet{Gerardi_Rosenblatt_Willen_Yao_2015} discuss the problem that if the treated ring is defined too narrowly, then control units will be affected by treatment causing a biased estimate of the counterfactual trend. \citet{Sullivan_2017} discusses the problem more formally and derives that the bias will be the difference in treatment effects experienced by the `treated' ring and the `control' ring. My paper expands on the results of \citet{Sullivan_2017} by including the additional source of bias that can result from a violation of parallel trends. Other researchers have recognized that estimating a single average treatment effect is less informative than a treatment effect curve. They solve this by using multiple rings to estimate treatment effects at different distances (e.g. \citet{Alexander_Currie_Schnell_2019,Casey_Schiman_Wachala_2018,Di_Tella_Schargrodsky_2004}). However, this approach selects multiple rings in an ad-hoc manner, still requires treatment effects to become zero after the outer-most treatment ring, and is prone to problems of specification searching. The current study's proposed estimator selects the number and location of rings in a data-driven way and does not require correct specification of where treatment effects become zero.

\citet{Diamond_McQuade_2019} propose a nonparametric estimator aimed at estimating a treatment effect surface. They use two-dimensions (latitude/longitude) to better approximate a smooth change in counterfactual outcomes (e.g. north-west and south-east from treatment might have different treatment effects). My method, instead, uses a singular measure of distance which pools units at similar distances but different directions from treatment and therefore delivers more precise estimates. However, the treatment effect estimate may mask heterogeneity of effects at different directions. If a researcher has a reason to suspect significant heterogeneity, then their proposed estimator will make a better fit.

This paper also contributes to a small literature on difference-in-differences estimators from a spatial lens \citep{Butts_2021,Clarke_2017,Berg_Streitz_2019,Verbitsky-Savitz_Raudenbush_2012,Delgado_Florax_2015}. These papers address instances where treatment is well defined by administrative boundaries but spillovers cause problems of defining who is `treated' and at what level of exposure. \citet{Butts_2021} and \citet{Clarke_2017} both recommend a method of using many rings to estimate treatment effects similar to the proposed nonparametric estimator. \citet{Clarke_2017} does specify a cross-validation approach for selecting rings, but does not specify that this result requires common local trends. Since this paper focuses on local shocks where constant parallel trends are plausible, I am able to provide a data-driven approach to choosing rings.

Last, there is a growing literature around design-based estimation of treatment effects in the presence of spillover effects \citep{Sävje_Aronow_Hudgens_2019,Aronow_Eckles_Samii_Zonszein_2020}. \citet{Aronow_Samii_Wang_2021} specifically discuss estimation of what this paper calls the treatment effect curve, or the average treatment effect at a certain distance away from treatment. This paper compliments this literature by introducing model-based assumptions for cases where treatment is not assigned following an experimental design.

\section{Example of Problem}\label{sec:example}

To illustrate the methodological difficulties in this method, I present an illustriative example. Suppose that an overgrown empty lot in a high-poverty neighborhood is cleaned up by the city and the outcome of interest is home prices. The researcher observes a panel of home sales before and after the lot is cleaned. Cleaning up the lot causes home values to go up directly nearby and as you move away from the lot, the positive treatment effect will decay to zero effect at, say, 3/4 of a mile. Since treatment is targeted to the high poverty neighborhood, comparisons with other neighborhoods in the cities could be biased if the neighborhood home prices are on different trends. Hence, the researcher wants to look only at the homes in the immediate neighborhood.


\begin{figure}[tb]
    \caption{Example of Problems with Ad-Hoc Ring Selection}
    \label{fig:problems}

    \begin{adjustbox}{width = 1.1\textwidth, center}
        \includegraphics[width=\textwidth]{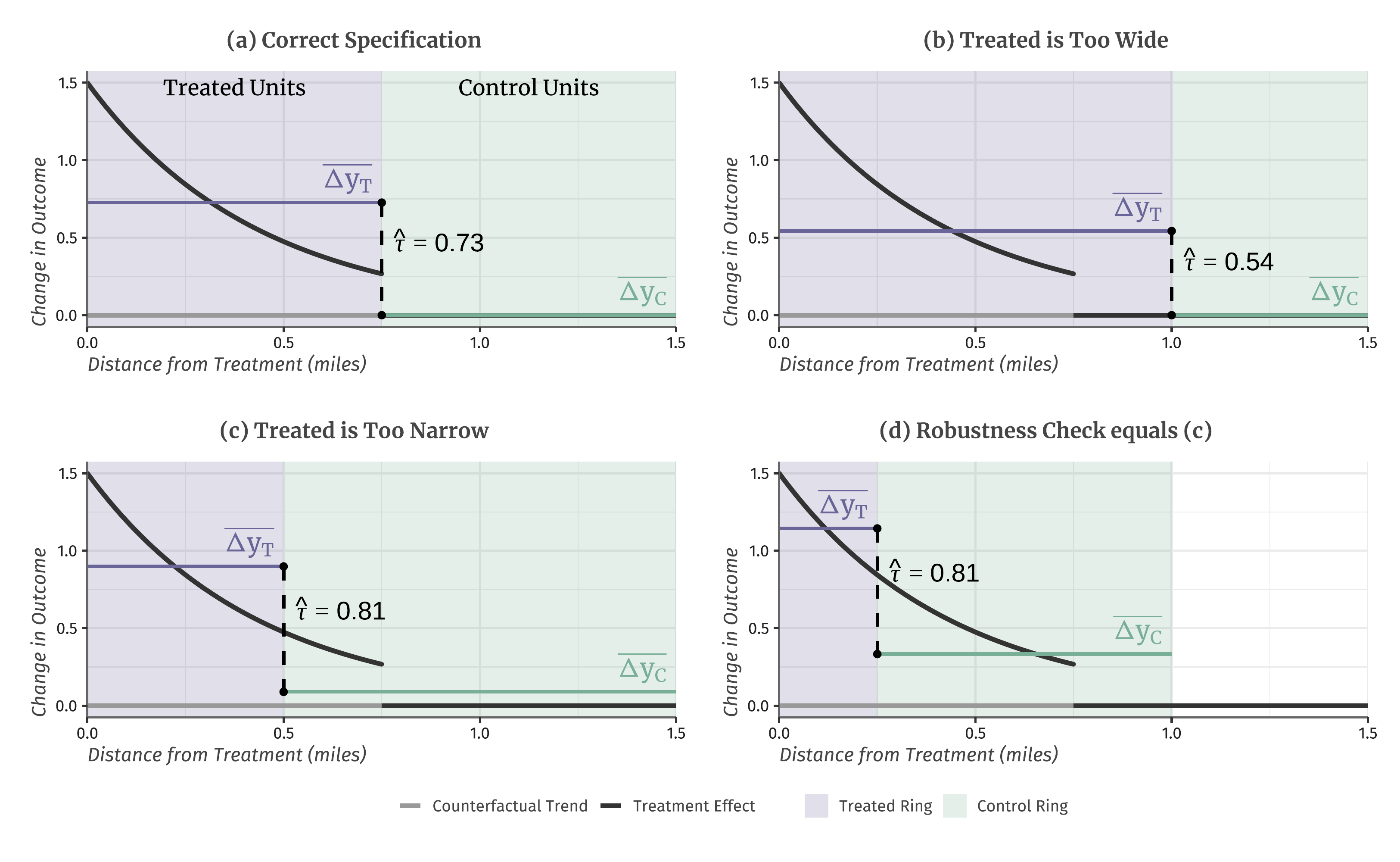}
    \end{adjustbox}

    {\footnotesize \emph{Notes:} This figure shows an example of the ring method. For each panel, the inner ring marks units considered `treated', the outer ring marks units considered `control' units, and the rest of the observations are removed from the sample. Then average changes in outcomes are compared between the treated and the control units to form a treatment effect estimate.}
\end{figure}

\autoref{fig:problems} shows a plot of simulated data from this example. The black line is treatment effect at different distances from the empty lot and the grey line is the underlying (constant) counterfactual change in home prices, normalized to 0. Panel (a) of \autoref{fig:problems} shows the best-case scenario where the treated ring is correctly specified. The two horizontal lines show the average change in outcome in the treated ring and the control ring. The treatment effect estimate, $\hat{\tau}$, is the difference between these two averages. However, this singular number masks over a large amount of treatment effect heterogeneity with units very close to treatment having a treatment effect double that of $\hat{\tau}$ and units near 3/4 miles experience a treatment efffect half as large as $\hat{\tau}$. For this reason, even if a researcher identifies the correct average treatment effect, they are masking a lot of heterogeneity that is potentially interesting. Therefore, later in this paper I recommend nonparametrically estimating the treatment effect curve as a function of distance rather than using average effect. 

However, the researcher does not typically know the distance at which treatment effects stop. Panels (b) and (c) highlights how treatment effect estimates change with a change in ring distances. Panel (b) shows when the `treatment' ring is too wide. In this case, some of the units in the treatment ring receive no effect from treatment and therefore makes the average treament effect among units in the treatment ring smaller. Therefore when the treated ring is too large, the estimated treatment effect is too small. Panel (c) of \autoref{fig:problems} shows the opposite case, where the treated ring is too narrow. In this case, there are some units in the `control' ring that experience treatment effects. Hence, the average change in outcome among the control unit is too large. This does not, though, decrease the treatment effect as one may suspect. Since the treatment effect decays with distance, the average change in outcome among the more narrow `treatment' ring is larger than the correct specification. The estimated treatment effect in this case grows, but it is not clear more generally whether the treatment effect will increase or decrease.\footnote{This primarily depends on the curvature of the treatment effect curve.} From these three examples, it's clear that the estimation strategy requires researchers to know the exact distance at which treatment effects become zero. Since this is a very demanding assumption, I propose an improved estimator in \autoref{sec:lspartition} that relaxes this assumption. 

Often times, researchers try multiple sets of rings and if the estimated effect remains similar across specifications, they assume the results are `robust'. Panel (d) of \autoref{fig:problems} shows an example of why this a problem. If Panel (c) was the researchers' original specification and Panel (d) was run as a robustness check, then the researcher would be quite confident in their results even though the estimate is too large in both cases. Now, I turn to econometric theory in order to formalize the intuition developed in this section.

\section{Theory}

Now, I develop econometric theory to formalize the intuition developed in the previous section. A researcher observes panel data of a random sample of units $i$ at times $t = 0, 1$ located in space at point $\theta_i = (x_i, y_i)$. Treatment occurs at a location $\bar{\theta} = (\bar{x}, \bar{y})$ between periods. Therefore, units differ in their distance to treatment, defined by $\dist_i \equiv d(\theta_i, \bar{\theta})$ for some distance metric $d$ (e.g. Euclidean distance) with a distribution function $F$. Outcomes are given by 
\begin{equation}
    Y_{it} = \mu_i + \tau_i \one_{t = 1} + \lambda_i \one_{t=1} +u_{it},    
\end{equation}
where $\mu_i$ is unit-specific time-invariant factors, $\lambda_i$ is the change in outcomes due to non-treatment shocks in period 1, $\tau_i$ is unit $i$'s treatment effect. Both $\lambda$ and $\tau$ can be split into a systematic function of distance $z(\dist_i)$ and an idiosyncratic term $\tilde{z}_i \equiv z_i - z(\dist_i)$ with $z$ being $\tau$ and $\lambda$. $\tau(d)$ is the average effect of treatment at a given distance and $\lambda(d)$ summarizes how covariates and shocks change over distance.
Therefore, we could rewrite our model as 
\begin{equation}\label{eq:model}
    Y_{it} = \mu_i + \tau(\dist_i) \one_{t = 1} + \lambda(\dist_i) \one_{t=1} + \varepsilon_{it},   
\end{equation}
where $\varepsilon = u_{it} + \tilde{\tau}_i + \tilde{\lambda}_i$ which is uncorrelated with distance to treatment. Researchers are trying to identify the average treatment effect on units experiencing treatment effects, i.e. $\bar{\tau} = \condexpec{\tau_i}{\tau(\dist_i) > 0}$.

\begin{assumption}[Random Sampling]
    The observed data consists of $\{ Y_{i1}, Y_{i0}, \dist_{i}\}$ which is independent and identically distributed.
\end{assumption}

Taking first-differences of our model, we have $\Delta Y_{it} = \tau(\dist_i) + \lambda(\dist_i) + \Delta \varepsilon_{it}$. It is clear that $\tau(\dist_i)$ and $\lambda(\dist_i)$ are not seperately identified unless additional assumptions are imposed. The central identifying assumption that researchers claim when using the ring method is that counterfactual trends likely evolve smoothly over distance, so that $\lambda(\dist_i)$ is approximately constant within a small distance from treatment. This is formalized in the context of our outcome model by the following assumption. 

\begin{assumption}[Local Parallel Trends]\label{assum:parallel}
    For a distance $\bar{d}$, we say that `local parallel trends' hold if for all positive $d, d' \leq \bar{d}$, then $\lambda(d) = \lambda(d')$.
\end{assumption}

This assumption requires that, in the absence of treatment, outcomes would evolve the same at every distance from treatment within a certain maximum distance, $\bar{d}$. To clarify the assumption, it is helpful to think of ways that it can fail. First, if treatment location is targeted based on trends \emph{within a small-area/neighborhood}, then trends would not be constant within the control ring. Second, if units sort either towards or away from treatment in a way that is systematically correlated with the outcome variable, then the compositional change can cause a violation in trends over time. Note that \nameref{assum:parallel} implies the standard assumption that parallel trends holds \emph{on average} between the treated and control rings:

\begin{assumption}[Average Parallel Trends]\label{assum:parallel_weak}
    For a pair of distances $d_t$ and $d_c$, we say that `average parallel trends' hold if $\condexpec{\lambda_d}{0 \leq d \leq d_t} = \condexpec{\lambda_d}{d_t < d \leq d_c}$.
\end{assumption}

If \nameref{assum:parallel} holds for some $d_c$, then our first-difference equation can be simplified to $\Delta Y_{it} = \tau(\dist_i) + \lambda + \Delta \varepsilon_{it}$ where $\lambda$ is some constant for units in the subsample $\mathcal{D} \equiv \{i \ : \ \dist_i \leq d_c \} $. Therefore, the treatment effect curve $\tau(\dist_i)$ is identifiable up to a constant under Assumption \ref{assum:parallel}. To identify $\tau(\dist_i)$ seperately from the constant, researchers will often claim that treatment effects stop occuring before some distance $d_t < d_c$. This is formalized in  the following assumption. 

\begin{assumption}[Correct $d_t$]\label{assum:dt}
    A distance $d_t$ satisfies this assumption if (i) for all $d \leq d_t$, $\tau(d) > 0$ and for all $d > d_t$, $\tau(d) = 0$ and (ii) $F(d_c) - F(d_t) > 0$.
\end{assumption}

With this assumption, the first difference equation simplifies to $\Delta Y_{it} = \lambda + \Delta \varepsilon_{it}$ for units with $d_t < \dist_i < d_c$. These units therefore identify $\lambda$. The `ring method' is the following procedure. Researchers select a pair of distances $d_t < d_c$ which define the ``treated'' and ``control'' groups. These groups are defined by $\mathcal{D}_t \equiv \{ i : 0 \leq \dist_i \leq d_t \}$ and $\mathcal{D}_c \equiv \{ i : d_t < \dist_i \leq d_c \}$. On the subsample of observations defined by $\mathcal{D} \equiv \mathcal{D}_t \cup \mathcal{D}_c$, they estimate the following regression:

\begin{equation}\label{eq:ring_method}
    \Delta Y_{it} = \beta_0 + \beta_1 \one_{i \in \mathcal{D}_t} + u_{it}.
\end{equation}

From standard results for regressions involving only indicators, $\hat{\beta}_1$ is the difference-in-differences estimator with the following expectation:
\[
    \expec{\hat{\beta}_1} = \condexpec{\Delta Y_{it}}{\mathcal{D}_t} - \condexpec{\Delta Y_{it}}{\mathcal{D}_c}.
\]
This estimate is decomposed in the following proposition.\footnote{A similar derivation of part (i) is found in \citet{Sullivan_2017} but does not include difference in parallel trends.}

\begin{proposition}[Decomposition of Ring Estimate]\label{prop:ring_decomp}  
    Given that units follow model (\ref{eq:model}),
    \begin{enumerate}
        \item[(i)] The estimate of $\beta_1$ in (\ref{eq:ring_method}) has the following expectation:
        \begin{align*}
            \expec{\hat{\beta}_1} &= \condexpec{\Delta Y_{it}}{\mathcal{D}_t} - \condexpec{\Delta Y_{it}}{\mathcal{D}_c} \\
            &=  \underbrace{\condexpec{\tau(\dist)}{\mathcal{D}_t} - \condexpec{\tau(\dist)}{\mathcal{D}_c} }_{\text{Difference in Treatment Effect}} + \underbrace{\condexpec{\lambda(\dist)}{\mathcal{D}_t} - \condexpec{\lambda(\dist)}{\mathcal{D}_c} }_{\text{Difference in Trends}}.
        \end{align*}
        
        \item[(ii)] If $d_c$ satisfies \nameref{assum:parallel} or, more weakly, if $d_t$ and $d_c$ satisfy \nameref{assum:parallel_weak}, then
        \[ 
            \expec{\hat{\beta}_1} = 
            \underbrace{\condexpec{\tau(\dist)}{\mathcal{D}_t} - \condexpec{\tau(\dist)}{\mathcal{D}_c} }_{\text{Difference in Treatment Effect}}.
        \] 
    
        \item[(iii)] If $d_c$ satisfies \nameref{assum:parallel} and $d_t$ satisfies Assumption \ref{assum:dt}, then
        \[ 
            \expec{\hat{\beta}_1} = \bar{\tau}.
        \]
    \end{enumerate}
\end{proposition}

Part (i) of this proposition shows that the estimate is the sum of two differences. The first difference is the difference in average treatment effect among units in the treated ring and units in the control ring. The second difference is the difference in counterfactual trends between the treated and control rings. This presents two possible problems. If some units in the control group experience effects from treatment, the average of these effects will be subtracted from the estimate. Second, since treatment can be targeted, the treated ring could be on a different trend than units further away and hence control units do not serve as a good counterfactual for treated units.

Part (ii) says that if $d_c$ satisfies \nameref{assum:parallel}, then the difference in trends from part (i) is equal to 0. As discussed above, the decomposition in part (ii) of Proposition \ref{prop:ring_decomp} is not necessarily unbiased estimate for $\bar{\tau}$. First, if $d_t$ is \emph{too wide}, then $\mathcal{D}_t$ contain units that are not affected by treatment. In this case, $\hat{\beta}_1$ will be biased towards zero from the inclusion of unaffected units from $d_t$ being too wide. Second, if $d_t$ is \emph{too narrow} then the $\mathcal{D}_c$ will contain units that experience treatment effects. It is not clear in this case, though, whether $\hat{\beta}_1$ will grow or shrink without knowledge of the $\tau(\dist)$ curve, but typically $\hat{\beta}_1$ will not be an unbiased estimate for $\bar{\tau}$. See the previous section for an example.  

Part (iii) of Proposition \ref{prop:ring_decomp} shows that if $d_t$ is correctly specified as the maximum distance that receives treatment effect, then $\hat{\beta}_1$ will be an unbiased estimate for the average treatment effect among the units affected by treatment. However, Assumption \ref{assum:dt} is a very demanding assumption and unlikely to be known by the researcher unless there are \emph{a priori} theory dictating $d_t$.\footnote{As an example, \citet{Currie_Davis_Greenstone_Walker_2015} uses results from scientific research on the maximum spread of local pollutants and \citet{Marcus_2021} use the plume length of petroleum smoke.} The following section will improve estimation by allowing consistent nonparametric estimation of the entire $\tau(\dist)$ function. An estimate of $\tau(\dist)$ can then be numerically integrated to for an estimate of $\bar{\tau}$.

\section{Nonparametric Estimation of the Treatment Effect Curve}\label{sec:lspartition}

In this section, I propose an estimation strategy that nonparametrically identifies the treatment effect curve $\tau(\dist_i)$ using partitioning-based least squares estimation and inference methods developed in \citet{Cattaneo_Crump_Farrell_Feng_2019, Cattaneo_Farrell_Feng_2019}. Partition-based estimators seperate the support of a covariate, $\dist_i$, into a set of quantile-spaced intervals (e.g. 0-25th percentiles of $\dist_i$, 25-50th, 50-75th, and 75-100th). Then the conditional $\condexpec{Y_i}{\dist_i}$ is estimated seperately within each interval as a $k$-degree polynomial of the covariate $x_i$.

For a given $d_c$, we will form a partition of our sample $\mathcal{D} = \{ i : \dist_i \leq d_c \}$ into $L$ intervals based on quantiles of the distance variable. Denote a given quantile as $\mathcal{D}_j \equiv\{ i : F_n^{-1}(\frac{j-1}{L}) \leq \dist_i < F_n^{-1}(\frac{j}{L}) \}$ where $F_n$ is the empirical distribution of $\dist$. Let $\{ \mathcal{D}_1, \dots, \mathcal{D}_L \}$ be the collection of the $L$ intervals. This paper will impose $k = 0$ which will predict $\Delta Y_{it}$ with a constant within each interval.\footnote{Approximation can be made arbitrarily close to the true conditional expectation function by \emph{either} increasing the number of intervals \emph{or} by increasing the polynomial order to infinity, so setting $k = 0$ does not impose any cost.} 

These averages are defined as 
\[
    \overline{\Delta Y}_j \equiv \frac{1}{n_j} \sum_{i \in \mathcal{D}_j} \Delta Y_{it},
\]
where the number of units in bin $\mathcal{D}_j$ is $n_j \approx n/L$. Our estimator for $\condexpec{\Delta Y_{it}}{\dist_i}$ is then given by
\[
    \widehat{\Delta Y_{it}} = \sum_{j = 1}^{L} \one_{i \in \mathcal{D}_j} \overline{\Delta Y}_j
\]
As the number of intervals approach infinity, this estimate will approach $\condexpec{\Delta Y_{it}}{\dist = d}$ in a mean-squared error sense. Under \nameref{assum:parallel}, $\condexpec{\Delta Y_{it}}{\dist = d} \equiv \condexpec{\tau(\dist)}{\dist = d} + \lambda$. To remove $\lambda$, we require a less-strict version of assumption \ref{assum:dt}.

\begin{assumption}[$d_t$ is within $d_c$]\label{assum:dt_weak}
    A distance $d_c$ satisfies this assumption if there exists a distance $d_t$ with $0 < d_t < d_c$ such that (i) Assumption \ref{assum:dt} holds and (ii) $F(d_c) - F(d_t) > 0$.
\end{assumption}

If a distance $d_c$ satisfies \nameref{assum:parallel} and (\ref{assum:dt_weak}), the mean within the last ring $\mathcal{D}_k$ will estimate $\lambda$ as the number of bins $L \to \infty$. The reason for this is simple, as $L \to \infty$, the last bin will have the left end-point $> d_t$ and therefore $\tau(\dist) = 0$ in $\mathcal{D}_L$. Under local parallel trends, the last ring will therefore estimate $\lambda$. Therefore, estimates of $\tau(\dist_i)$ can be formed for each interval as $\hat{\tau}_j \equiv \overline{\Delta Y}_j - \overline{\Delta Y}_L$. 

\begin{proposition}[Consistency of Nonparametric Estimator]\label{prop:np_identification}  
    Given that units follow model (\ref{eq:model}) and $d_c$ satisfies \nameref{assum:parallel} and assumption (\ref{assum:dt_weak}), as $n$ and $L \to \infty$ 

    \begin{align*}
        \hat{\tau} &\equiv \sum_{i=1}^L \hat{\tau}_j 1_{i \in \mathcal{D}_j} \to^{unif} \tau(\dist)
    \end{align*}

    where $d_j$ corresponds to the $F^{-1}(D_j)$.
\end{proposition}

As discussed in Section \ref{sec:example}, specifying $d_t$ correctly is important to identify the average treatment effect among the affected in the parametric estimator. The nonparametric estimator only requires that treatment effects become zero before $d_c$, i.e. that such a $d_t$ exists. However, the estimator would no longer identify the treatment effect curve under the milder \nameref{assum:parallel_weak} assumption. Therefore, a researcher should justify explicity the assumption that, within the $d_c$ ring, every unit is subject to the same trend. This is most likely to be satisfied on a very local level and not very plausible in the case of larger units, e.g. counties.

The nonparametric approach allows estimation of the treatment effect curve whereas the indicator approach, \emph{at best}, can only estimate an \emph{average} effect among units experiencing effects. The treatment effect curve allows researcher to understand differences in treatment effect across distance. For example, typically one would assume treatment effects shrink over distance and evidence of this from the nonparametric approach can strengthen a causal claim. In some cases, such as a negative hyper-local shock and a postivie local shock (e.g. a local bus-stop), the treatment effect can even change sign across distances. In this case, the average effect could be near zero even though there are significant effects occuring. 

Plotting estimates $\hat{\tau}_j$ can provide visual evidence for the underlying \nameref{assum:parallel} assumption. Typically, treatment effect will stop being experienced far enough away from $d_c$ that some estimates of $\hat{\tau}_j$ with $j$ `close to' $L$ will provide informal tests for parallel trends holding. \autoref{fig:lr} provide an example where plotting of $\hat{\tau}_j$ provide strong evidence in support of local parallel trends as it appears that after some distance, average effects are consistetly centered around zero. This is not a formal test as it could be the case that the true treatment effect curve, $\tau(\dist)$ is perfectly cancelling out with the counterfactual trends curve $\lambda(\dist)$ producing near zero estimates, but this is a knive's edge case. 

The above proposition shows that the series estimator will consistenly estimate the treatment effect curve, $\tau(\dist)$ as the number of bins $L$ and the number of observations $n$ both go to infinity. In finite-samples though, we will have a fixed $L$ and hence a fixed set of treatment effect estimates $\{ \tau_1, \dots, \tau_{L}\}$ with $\tau_L \equiv 0$ by definition. The estimates $\hat{\tau}_j$ are approximately equal to $\condexpec{\tau(\dist)}{\dist \in \mathcal{D}_j}$ or the average treatment effect within the interval $\mathcal{D}_j$.

The choice of $L$ in finite samples is not entirely clear. \citet{Cattaneo_Crump_Farrell_Feng_2019} derive the IMSE-optimal choice of $L$ which is a completely data-driven choice. The optimal $L$ is driven by two competing terms in the IMSE formula. On the one hand, as $L$ increases, the conditional expectation function is allowed to vary more across values of $\dist$ and hence bias of the estimator decreases. However, larger values of $L$ increase the variance of the estimator. Balancing this trade-off depends on the shape and curvature of $\tau(\dist)$. The resulting choice of $L^*$ and the use of quantiles of the data allows a completely data-driven choice of the number of rings and their endpoints which allows for estimation in a principled and objective way. This principled estimator removes researcher-incentives to search across choices of rings to provide the best evidence. 

For a given $L^*$, \citet{Cattaneo_Crump_Farrell_Feng_2019} show the large-sample asymptotics of the estimates $\overline{\Delta Y}_j$ and provide robust standard errors for the conditional means that account for the additional randomness due to quantile estimation. Since our estimator is a difference in means, standard errors on our estimate $\hat{\tau}_j$ are given by $\sqrt{\sigma^2_j + \sigma^2_L}$, where $\sigma_j$ is the standard error recommended by \citet{Cattaneo_Crump_Farrell_Feng_2019}. These standard errors are produced by the Stata/R package \texttt{binsreg}. Inference can be done by using the estimated t-stat with the standard normal distribution. There may be concerned that the standard errors need to adjust for spatial correlation. However, this is not the case under assumption (\ref{assum:parallel}) as this implies the error term is uncorrelated with distance.

\section{Application to Neighborhood Effects of Crime Risk}

To highlight the advantages of my proposed estimator, I revisit the analysis of \citet{Linden_Rockoff_2008}. This paper analyzes the effect of a sex offender moving to a neighborhood on home prices. This paper uses the ring method with treated homes being defined as being within $1/10^{th}$ of the sex offender's home and the control units being between $1/10^{th}$ and $1/3^{rd}$ of a mile from the home. The authors make a case for the ring method by arguing that \emph{within a neighborhood}, \nameref{assum:parallel} holds since they are looking in such a narrow area and purchasing a home is difficult to be precisely located with concurrent hyper-local shocks.

\begin{figure}[htb!]
    \caption{Price Gradient of Distance from Offender}\label{fig:lr_nonparametric}

    \begin{subfigure}{0.33\textwidth}
        \caption{Bandwidth of 0.025}
    \end{subfigure}
    \begin{subfigure}{0.33\textwidth}
        \caption{Bandwidth of 0.075}
    \end{subfigure}
    \begin{subfigure}{0.33\textwidth}
        \caption{Bandwidth of 0.125}
    \end{subfigure}
    
    \vspace{-3mm}
    \includegraphics[width=\textwidth]{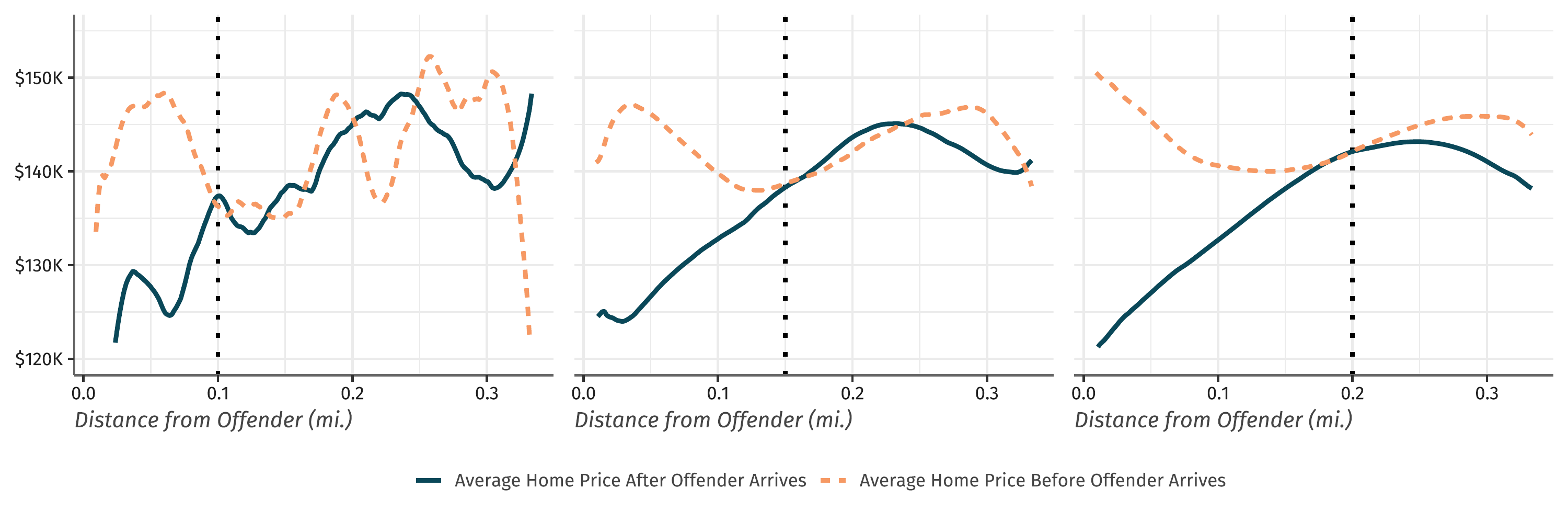}

    {\footnotesize{\it Notes:} This figure plots estimates of home prices in the year before and the year after the arrival of a sex offender estimated using a Local Polynomial Kernel Density estimation with an Epanechnikov kernel. Panel (b) recreates Figure 2 from \citet{Linden_Rockoff_2008} and the other panels change the bandwidth.}
\end{figure}

As for the choice of the treatment ring, there is little \emph{a priori} reasons to know how far the effects of sex offender arrival will extend in the neighborhood. The authors provide graphical evidence of nonparametric estimates of the conditional mean home price at different distances in the year before and the year after the arrival of a sex offender. The published plot can be seen in Panel (b) of \autoref{fig:lr_nonparametric}. They `eyeball' the point at which the two estimates are approximately equal to decide how far treatment effects extend. However, this approach is less precise than it may seem. Panels (a) and (c) show that changing the bandwidth for the kernel density estimator will produce very different guesses at how far treatment effects extend. My proposed estimator works in a data-driven way that does not require these ad-hoc decisions.

\begin{figure}[tb!]
    \caption{Effects of Offender Arrival on Home Prices \citep{Linden_Rockoff_2008}}\label{fig:lr}

    \begin{subfigure}{\textwidth}
        \caption{Indicator Approach}
        \centering
        \vspace{-2.5mm}
        \includegraphics[width=\textwidth]{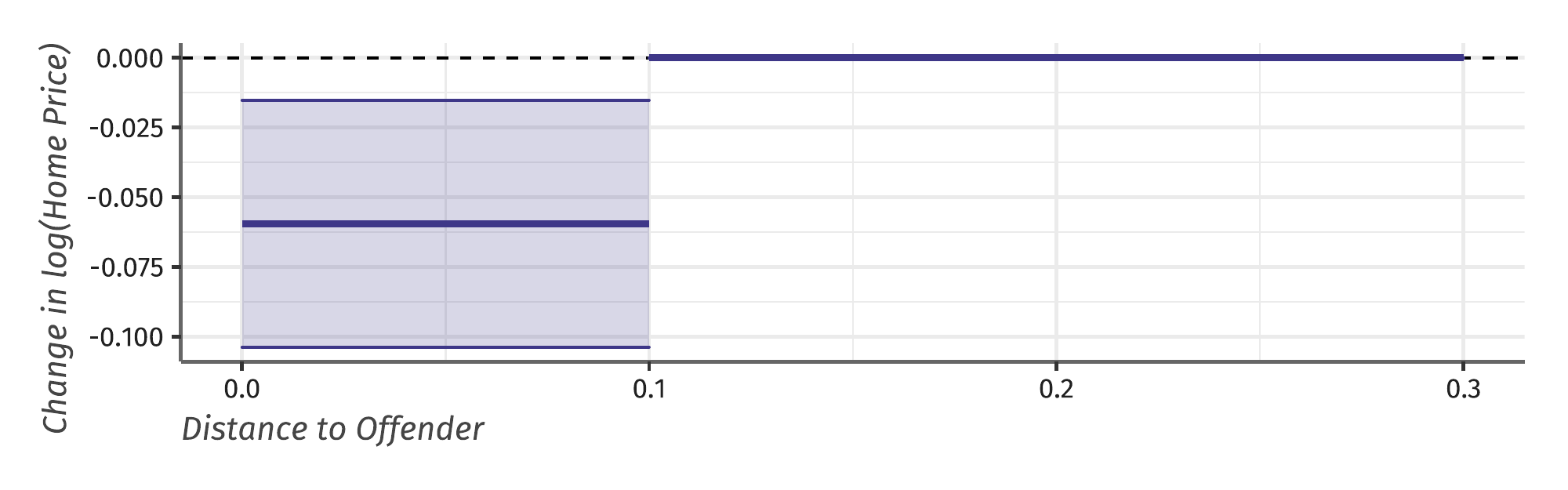}
    \end{subfigure}
    \hfill
    \begin{subfigure}{\textwidth}
        \caption{Nonparametric Approach}
        \centering
        \vspace{-2.5mm}
        \includegraphics[width=\textwidth]{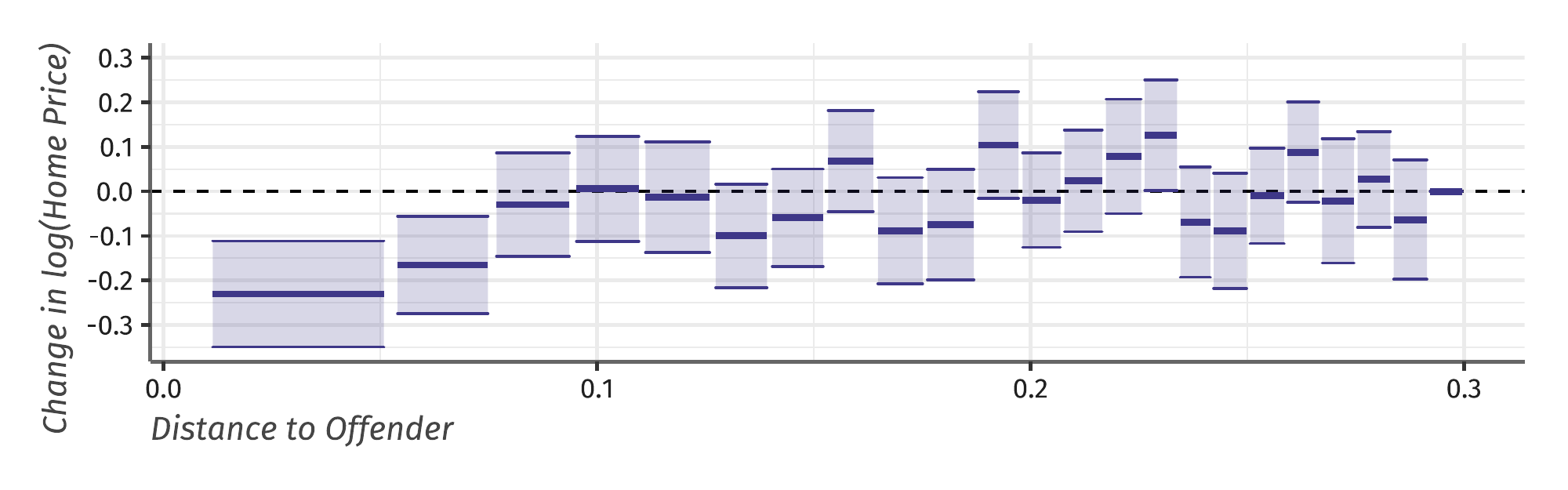}
    \end{subfigure}

    {\footnotesize{\it Notes:} This figure plots the estimated change in home prices after the arrival of a registered sex offender as a function of distance from offender. Each line plots $\hat{\tau}_j = \overline{\Delta Y}_j - \overline{\Delta Y}_l $ with associated standard errors. Panel (a) shows an estimate from Equation \ref{eq:ring_method} with a treatment distance of $1/10^{th}$ miles and a control distance of $1/3^{rd}$ mile. Panel (b) shows the nonparametric estimate of $\tau(\dist_i)$ proposed in Section \ref{sec:lspartition}.}
\end{figure}

The standard rings approach is equivalent to my proposed method with two rings: $\mathcal{D}_1$ being the treated homes between 0 and 0.1 miles away and $\mathcal{D}_2$ being the control homes between 0.1 and 0.3 miles away. The average change among $\mathcal{D}_2$ estimates the counterfactual trend and the average change among $\mathcal{D}_1$ minus the estimated counterfactual trend serves as the treatment effect. Panel (a) of \autoref{fig:lr} shows the basic results of their difference-in-differences analysis which plots estimates $\hat{\tau}_j$ for $j = 1,2$.  On average, homes between 0 and 0.1 miles decline in value by about 7.5\% after the arrival of a sex offender. \emph{As an assumption} of the rings method, homes between 0.1 and 0.3 miles away are not affected by a sex offender arrival. The choice of 0.1 miles is an untestable assumption and as seen above the evidence provided is highly dependent on the choice of bandwidth parameter. My proposed estimator does not require a specific choice for a `treated' area.

\citet{Linden_Rockoff_2008} only have access to a non-panel sample of home sales, so identification requires another assumption for identification, namely that the composition of homes at a given distance does not change over time. Further, since we can no longer form first-differences of the outcome variable, seperate nonparametric estimators must be estimated before and after treatment and subtracted from one another. Details of this theory are in the Online Appendix

Panel (b) of \autoref{fig:lr} applies the nonparametric approach described in Section \ref{sec:lspartition}. Two differences in results occur. First, homes in the two closest rings i.e. within a few hundred feet, are most affected by sex-offender arrival with an estimated decline of home value of around 20\%. homes a bit further away but still within in Linden and Rockoff's `treated' sample do not experience statistically significant treatment effects. As discussed above, Linden and Rockoff's estimate of $\bar{\tau}$ is attenuated towards zero because of the inclusion of homes with little to no treatment effects, leading them to understate the effect of arrival on home prices. The nonparametric approach improves on answering this question by providing a more complete picture of the treatment effect curve. The magnitude of treatment effects decrease over distance, providing additional evidence that the arrival causes a drop in home prices.\footnote{This is similar to estimating a dose-response function as evidence supporting a causal mechanism.} 

The second advantage of this approach is that the produced figure provides an informal test of the local parallel trends assumption. After 0.1 miles, the estimated treatment effect curve becomes centered at zero consistently. This implies that units within each ring have the same estimated trend as the outer most ring, providing suggestive evidence that homes in this neighborhood are subject to the same trends.

\section{Discussion}

This article formalizes a common applied identification strategy that has a strong intuitive appeal. When treatment effects of shocks are experienced in only part of an area that would otherwise be on a common neighborhood-trend, difference-in-differences comparisons within a neighborhood can identify treatment effects. However, this paper shows that the typical \emph{estimator} for treatment effects requires a very strong assumption and returns only an average treatment effect among affected units when this assumption holds. 

This article then proposes an improved estimator that relies on nonparametric series estimators. The nonparametric estimator allows for estimation of the treatment effect at different distances from treatment, similar to a dose-response function, which can allow better understanding of \emph{who} is experiencing effects and how this changes across `exposure' to a shock. More, in some cases it can provide explanation for null results. For example, if a bus station creates negative externalities for apartments that border the station but positive externalities for apartments within walking distance, the average effect could be zero. However, nonparametric estimation would reveal the two effects seperately.

\newpage~\bibliography{references.bib}

\appendix 

\section{Proofs}
\label{sec:proofs}

\subsection{Proof of Proposition \ref{prop:ring_decomp}}

\begin{proof}
    \ Note using our model (\ref{eq:model}), we have 
    \begin{align*}
        \expec{\hat{\beta}_1} &= \condexpec{\Delta Y_{it}}{\mathcal{D}_t} - \condexpec{\Delta Y_{it}}{\mathcal{D}_c} \\
        &= \condexpec{\tau(\dist_i) + \lambda(\dist_i) + \Delta \varepsilon_{i}}{\mathcal{D}_t} - \condexpec{\tau(\dist_i) + \lambda(\dist_i) + \Delta \varepsilon_{i}}{\mathcal{D}_c} \\ 
        &= \condexpec{\tau(\dist_i)}{\mathcal{D}_t} - \condexpec{\tau(\dist_i)}{\mathcal{D}_c} + \condexpec{\lambda(\dist_i)}{\mathcal{D}_t} - \condexpec{\lambda(\dist_i)}{\mathcal{D}_c} + \condexpec{\Delta\varepsilon_i}{\mathcal{D}_t} - \condexpec{\Delta\varepsilon_i}{\mathcal{D}_c}.
    \end{align*}
    By construction, $\Delta \varepsilon_i$ is uncorrelated with distance, so the final two terms in the sum is zero giving us result (i). Result (ii) comes from the fact that within $d_c$, $\lambda(\dist_i) = \lambda$. Result (iii) comes from the fact that if $d_t$ is the correct cutoff $\condexpec{\tau(\dist_i)}{\mathcal{D}_c} = 0$.
\end{proof}

\subsection{Proof of Proposition \ref{prop:np_identification}}

\begin{proof}
    \ Note that $L \to \infty$ implies $d_t \leq F_n^{-1}(\frac{L-1}{L})$ by assumption (\ref{assum:dt_weak}). This implies $\overline{\Delta Y}_L \to^p \lambda$ as $n \to \infty$ by assumption (\ref{assum:dt_weak}). 
    
    From assumption (\ref{assum:parallel}) and from our model (\ref{eq:model}), we have
    \begin{align*}
        \hat{\tau}_j &= \overline{\Delta Y}_j - \overline{\Delta Y}_L \\
        &\to^p \condexpec{\tau(\dist)}{\dist \in \mathcal{D}_j} + \lambda - \lambda \\
        &= \condexpec{\tau(\dist)}{\dist \in \mathcal{D}_j}
    \end{align*}

    As $L \to \infty$ and $n \to \infty$, we have that $\mathcal{D}_j$ approaches a set containing a singular point, say $d_j$. Therefore 
    $$ 
        \hat{\tau}_j \to^p \condexpec{\tau(\dist)}{\dist = d_j}
    $$

    The sum of $\hat{\tau}_j$ therefore approach the conditional expectation function of $\tau(\dist)$ pointwisely. See \citet{Cattaneo_Farrell_Feng_2019} for proof of uniform convergence and underlying smoothness conditions for nonparametric consistency.
\end{proof}

\section{Repeated Cross-Sectional Data}

In the case of repeated cross-sectional data, we have individuals $i$ that appear in the data in period $t(i) \in \{0,1\}$. However, since we no longer are able to observe units in both periods, we are not able to take first differences. Our model therefore will have a $\lambda$ term for both periods. Therefore, $\lambda$ includes the average of $\mu_i$, covariates, and period shocks at a given distance.

\begin{equation}\label{eq:model_rc}
    Y_{i} = \tau(\dist_i) \one_{t(i) = 1} + \lambda_{t(i)}(\dist_i) + \nu_{it}.  
\end{equation}

The parallel trends assumption must be modified now in the case of cross sections:

\begin{assumption}[Local Parallel Trends (RC)]\label{assum:parallel_rc}
    For a distance $\bar{d}$, we say that `local parallel trends' hold if for all positive $d, d' \leq \bar{d}$, then $\lambda_1(d) - \lambda_0(d) = \lambda_1(d') - \lambda_0(d')$.
\end{assumption}

\begin{assumption}[Average Parallel Trends (RC)]\label{assum:parallel_weak_rc}
    For a pair of distances $d_t$ and $d_c$, we say that `average parallel trends' hold if $\condexpec{\lambda_1(d)}{0 \leq d \leq d_t} - \condexpec{\lambda_0(d)}{0 \leq d \leq d_t} = \condexpec{\lambda_1(d)}{d_t < d \leq d_c} - \condexpec{\lambda_0(d)}{d_t < d \leq d_c}$.
\end{assumption}

The parallel trends assumption is a bit more complicated now and is theoretically stricter. \nameref{assum:parallel_rc} still require changes in outcomes over time for a given unit $i$ must be constant across distance (or on average in the case of \nameref{assum:parallel_weak_rc}). However since the composition of units can change over time, this also requires that the average of individual fixed effects must be constant across time. This is well understood in the hedonic pricing literature that the composition of homes being sold can not change over time for identification (e.g. \citet{Linden_Rockoff_2008}).

For completeness, I rewrite the other necessary assumption
\begin{assumption}[Correct $d_t$]\label{assum:dt_repeated}
    A distance $d_t$ satisfies this assumption if (i) for all $d \leq d_t$, $\tau(d) > 0$ and for all $d > d_t$, $\tau(d) = 0$ and (ii) $F(d_c) - F(d_t) > 0$.
\end{assumption}

The ring estimate in the case of cross-sections is given by:
\begin{equation}\label{eq:ring_method_repeated}
    \Delta Y_{i} = \beta_0 + \beta_1 \one_{i \in \mathcal{D}_c} \one_{t(i) = 1} + \beta_2 \one_{i \in \mathcal{D}_t} \one_{t(i) = 0} + \beta_4 \one_{i \in \mathcal{D}_t} \one_{t(i) = 1} + u_{it}.
\end{equation}

\begin{proposition}[Decomposition of Ring Estimate (RC)]\label{prop:ring_decomp_rc}  
    Given that units follow model (\ref{eq:model_rc}),
    \begin{enumerate}
        \item[(i)] The estimate of $\beta_4$ in (\ref{eq:ring_method_repeated}) has the following expectation:
        \begin{align*}
            \expec{\hat{\beta}_4} &= \condexpec{\Delta Y_{it}}{\mathcal{D}_t} - \condexpec{\Delta Y_{it}}{\mathcal{D}_c} \\
            &=  \underbrace{\condexpec{\tau(\dist)}{\mathcal{D}_t} - \condexpec{\tau(\dist)}{\mathcal{D}_c} }_{\text{Difference in Treatment Effect}} \\
            &\quad + ( \condexpec{\lambda_1(\dist)}{\mathcal{D}_t, t(i) = 1} - \condexpec{\lambda_0(\dist)}{\mathcal{D}_t, t(i) = 0} ) \\
            &\quad - ( \condexpec{\lambda_1(\dist)}{\mathcal{D}_c, t(i) = 1} - \condexpec{\lambda_0(\dist)}{\mathcal{D}_c, t(i) = 0} )   \\
        \end{align*}
        
        \item[(ii)] If $d_c$ satisfies \nameref{assum:parallel_rc} or, more weakly, if $d_t$ and $d_c$ satisfy \nameref{assum:parallel_weak_rc}, then
        \[ 
            \expec{\hat{\beta}_4} = 
            \underbrace{\condexpec{\tau(\dist)}{\mathcal{D}_t} - \condexpec{\tau(\dist)}{\mathcal{D}_c} }_{\text{Difference in Treatment Effect}}.
        \] 
    
        \item[(iii)] If $d_c$ satisfies \nameref{assum:parallel_rc} and $d_t$ satisfies Assumption \ref{assum:dt_repeated}, then
        \[ 
            \expec{\hat{\beta}_4} = \bar{\tau}.
        \]
    \end{enumerate}
\end{proposition}

\begin{proof}
    \ With some algebraic manipulation, we can rewrite our difference-in-differences estimator as
    \begin{align*}
        &\left(\condexpec{Y_{i}}{\mathcal{D}_t, t(i) = 1} - \condexpec{Y_{i}}{\mathcal{D}_t, t(i) = 0}\right) - \left(\condexpec{Y_{i}}{\mathcal{D}_c, t(i) = 1} - \condexpec{Y_{i}}{\mathcal{D}_c, t(i) = 0})\right) \\
        &\quad = \left(\condexpec{ \tau(\dist_i) + \lambda_1(\dist_i) + \nu_{i1}}{\mathcal{D}_t, t(i) = 1} - \condexpec{\lambda_0(\dist_i) + \nu_{i0}}{\mathcal{D}_t, t(i) = 0}\right) \\
        &\quad\quad - \left(\condexpec{ \tau(\dist_i) + \lambda_1(\dist_i) + \nu_{i1}}{\mathcal{D}_c, t(i) = 1} - \condexpec{\lambda_0(\dist_i) + \nu_{i0}}{\mathcal{D}_c, t(i) = 0}\right) \\
        &\quad = \condexpec{\tau(\dist_i)}{\mathcal{D}_t, t(i) = 1} - \condexpec{\tau(\dist_i)}{\mathcal{D}_c, t(i) = 0} + \\
        &\quad\quad \left( \condexpec{\lambda_1(\dist_i)}{\mathcal{D}_t, t(i) = 1} - \condexpec{\lambda_0(\dist_i)}{\mathcal{D}_t, t(i) = 0} \right) \\
        &\quad\quad - \left( \condexpec{\lambda_1(\dist_i)}{\mathcal{D}_c, t(i) = 1} - \condexpec{\lambda_0(\dist_i)}{\mathcal{D}_c, t(i) = 0} \right),
    \end{align*}
    where the terms consisting of $\nu_{it}$ cancel out as they are uncorrelated with distance. Propositions (ii) and (iii) follow the same arguments as in the panel case.
\end{proof}

Part (i) of this theorem shows that under no parallel trends assumption, the `Difference in Trends' term becomes the change in $\lambda_t$ for the treated ring minus the change in for the control ring. As discussed above, this change in lambdas can be do to period specific shocks or changes in the composition of units observed in each period.

\subsection{Nonparametric Estimation}

Since we can no longer perform a single nonparmaetric regression on first differences in the context of cross-sections, our nonparametric estimator must be adjusted. The modified procedure will fit a nonparametric estimate of $\condexpec{Y_i}{\dist_i, t}$ seperately for $t = 0$ and $t = 1$ with a restriction that the bin intervals $\{ \mathcal{D}_1, \dots, \mathcal{D}_{L} \}$ must be the same in both samples.\footnote{The number of  intervals are decided based on a different IMSE condition described in \citet{Cattaneo_Crump_Farrell_Feng_2019} and the quantiles are calculated using the distribution of distances in both periods.} Then, for each distance bin we calculate an estimate of $\bar{Y}_{j,t}$ which corresponds to the sample average of observations in period $t$ in bin $\mathcal{D}_j$. 

Then estimates of $\tau_j$ can be formed as
\[
    \hat{\tau}_j = \left[\bar{Y}_{j,1} - \bar{Y}_{j,0}\right] - \left[\bar{Y}_{L,1} - \bar{Y}_{L,0}\right],
\]
where, as before, the change in trends in the last ring serve as an estimate for the counterfactual trend. Under \nameref{assum:parallel_rc}, estimates of $\hat{tau}_j$ are consistent for $\condexpec{\tau(\dist_i)}{i \in \mathcal{D}_j}$ and the treatment effect curve converges uniformly to the treatment effect curve $\tau(d)$. 

Standard errors are formed similarly as before, but is the difference of four means so they can be formed as $\sqrt{\sigma_{j,1}^2 + \sigma_{j,0}^2 + \sigma_{L,1}^2 + \sigma_{L,0}^2}$. These individual estimates and standard errors can be produced by the Stata/R package \texttt{binsreg}.

\end{document}